\documentclass[a4paper]{article}

\usepackage[english]{babel}
\usepackage[utf8x]{inputenc}
\usepackage[T1]{fontenc}
\usepackage{array}
\usepackage[a4paper,top=3cm,bottom=2cm,left=3cm,right=3cm,marginparwidth=1.75cm]{geometry}
\usepackage{amsmath}
\usepackage{graphicx}
\usepackage[colorinlistoftodos]{todonotes}
\usepackage[colorlinks=true, allcolors=blue]{hyperref}

\begin{document}

\title{Non-linear optical spectroscopy and two-photon excited fluorescence spectroscopy reveal the excited states of fluorophores embedded in beetle's elytra}

\maketitle

{\bf S\'{e}bastien R. Mouchet}$^{1,2,\dag,*}$, {\bf Charlotte Verstraete}$^{3,\dag}$, {\bf Dimitrije Mara}$^{4}$, {\bf Stijn Van Cleuvenbergen}$^{3}$, {\bf Ewan D. Finlayson}$^{1}$, {\bf Rik Van Deun}$^{4}$, {\bf Olivier Deparis}$^{2}$, {\bf Thierry Verbiest}$^{3}$, {\bf Bjorn Maes}$^{5}$, {\bf Pete Vukusic}$^{1}$, {\bf and Branko Kolaric}$^{5,6,**}$\\
$^{1}$School of Physics, University of Exeter, Stocker Road, Exeter EX4 4QL, United Kingdom\\
$^{2}$Department of Physics, University of Namur, Rue de Bruxelles 61, 5000 Namur, Belgium\\
$^{3}$Molecular Imaging and Photonics, Department of Chemistry, KU Leuven, Celestijnenlaan 200D, 3001 Heverlee, Belgium\\
$^{4}$L$^{3}$ \textendash~Luminescent Lanthanide Lab, Department of Inorganic and Physical Chemistry, Ghent University, Krijgslaan 281-S3, 9000 Ghent, Belgium\\
$^{5}$Micro- and Nanophotonic Materials Group, University of Mons, Place du Parc 20, 7000 Mons, Belgium\\
$^{6}$Department of Materials Physics and Physical Chemistry, Institute of General and Physical Chemistry, Studentski trg 12/V, 11000, Belgrade, Serbia\\

$^{\dag}$Co-shared first authorship\\
$^{*}$s.mouchet@exeter.ac.uk
$^{**}$branko.kolaric@umons.ac.be

\begin{abstract}
Upon illumination by ultraviolet light, many animal species emit light through fluorescence processes arising from fluorophores embedded within their biological tissues. Fluorescence studies in living organisms are however relatively scarce and so far limited to the linear regime. Multiphoton excitation fluorescence analyses as well as non-linear optical techniques offer unique possibilities to investigate the effects of the local environment on the excited states of fluorophores. Herein these techniques are applied for the first time to the study of insects’ natural fluorescence. The case of the male \textit{Hoplia coerulea} beetle is investigated because the scales covering the beetle's elytra are known to possess an internal photonic structure with embedded fluorophores, which controls both the beetle's colouration and the fluorescence emission. An intense two-photon excitation fluorescence signal is observed, the intensity of which changes upon contact with water. A Third-Harmonic Generation signal is also detected, the intensity of which depends on the light polarisation state. The analysis of these non-linear optical and fluorescent responses unveils the multi-excited states character of the fluorophore molecules embedded in the beetle's elytra. The anisotropy of the photonic structure, which causes additional tailoring of the beetle's optical responses, is confirmed by circularly polarised light and non-linear optical measurements.
\end{abstract}

Fluorescence occurs in many living organisms ranging from insects and birds to amphibians, mammals and plants~\cite{Pavan1954,Tani2004,Vukusic2005,Welch2012,Lagorio2015,Marshall2017,Taboada2017}. These organisms' tissues emit longer-wavelength light (typically in the visible band) when they are illuminated by shorter-wavelength light (usually, blue or ultraviolet). Such a light emission process is due to embedded fluorescent pigments, called fluorophores, such as green fluorescent protein (GFP), psittacofulvin, biopterin and papiliochrome II, within the biological integuments of the living organisms. It results from the transition of fluorophores' electrons between real states with the same spin multiplicity. The colours of emitted light span the whole visible electromagnetic spectrum from blue and green to yellow, orange and red. This phenomenon is thought to play a role in some living organisms' visual communication such as species recognition, mate choice, agonistic behaviour, prey detection and camouflage~\cite{Lagorio2015,Marshall2017}. Despite these important roles, fluorescence emission still remains under-explored from the biological, chemical and physical points of view, specifically in insects. Of course, in many cases, discerning whether fluorescence emission plays a functional role in the species' colouration and visual pattern or whether it is a non-functional consequence of the presence of specific molecules within the living organism's tissues may be complicated. To date, only a handful of organisms have been demonstrated to display fluorescence emission associated with biological functions~\cite{Lagorio2015,Marshall2017}.

\begin{figure}
\includegraphics[width=\columnwidth]{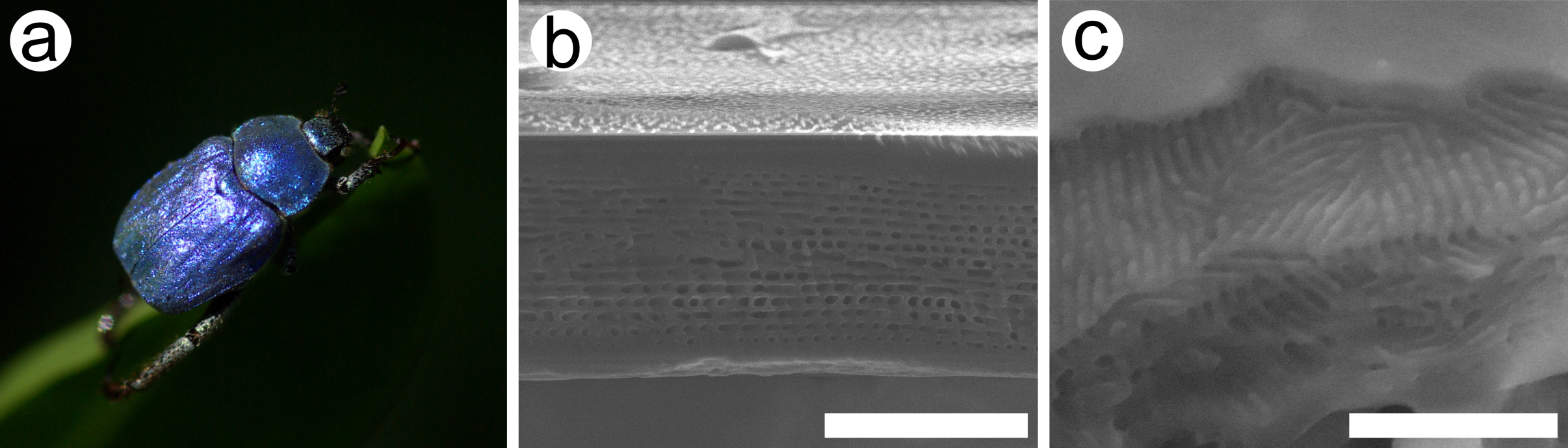}
\caption{\textbf{Structurally-coloured elytra of \textit{H. coerulea}.} (a) The blue-violet iridescent colour displayed by the male \textit{H. coerulea} beetle is due to (b) the presence of a periodic porous multilayer within the scales covering the beetle's elytra. This structure comprises thin layers made of cuticle material and mixed air-cuticle layers. (c) The latter exhibit form anisotropy due to rod-like spacers comprised by these layers. Details regarding the morphological characterisation can be found in Supplementary Information~1. Scale bars: (b,c) 2~µm.}
\label{fig:panel1}
\end{figure}

The male \textit{Hoplia coerulea} beetle (Figure~\ref{fig:panel1}a), investigated here, is known to display an iridescent blue-violet colouration due to a porous periodic multilayer made of chitin found in the scales covering its body (Figure~\ref{fig:panel1}b)~\cite{Vigneron2005}. Upon contact with liquids and vapours, its colour was found to turn green~\cite{Rassart2009,Mouchet2016b,Mouchet2016a,Mouchet2017b} due to the penetration of liquids into the scales. Interestingly, this photonic structure was demonstrated to contain naturally embedded fluorophores~\cite{VanHooijdonk2012,Mouchet2016c,Mouchet2017a} and to give rise to controlled fluorescence emission from these fluorophores, through modifications of the optical system's local density of optical states~\cite{Mouchet2016c,Mouchet2017a,Kolaric2007,Gonzalez2012}. Upon contact with water, the fluorescence emission caused by one-photon absorption from these confined fluorophores was observed to change from turquoise to navy blue~\cite{Mouchet2016c}. These water-induced changes were found to be reversible, implying that the fluorophores were not significantly altered from a chemical point of view by exposure to water.

This male beetle reflects incident light specularly. Fluorescence emission from its elytra takes place in a much larger solid angle than the angle of the very directional light reflection and could play a role in the signalling behaviour of the species. However, this species could have simply evolved a diffusively scattering surface that would be more efficient and scatter more light than the light emitted by fluorescence due to the known poor conversion efficiency of such processes in natural fluorophores, and the fact that sun light is less intense in the UV range with respect to the visible range. The role of these fluorophores might simply be a by-product of the photoprotection function due to the increased UV absorption induced by the presence of these molecules, leading to a better protection of the beetle's DNA.

The chemical and physical aspects of this fluorescent process are far from understood. Fluorescence emission from natural organisms' photonic structures has never been investigated, neither by multiphoton excitation fluorescence techniques, nor by non-linear optical techniques, although both are known as invaluable tools for the imaging and spectroscopy analyses of biological tissues~\cite{Rowlands2017}. Using One-Photon Excited Fluorescence (OPEF) and Two-Photon Excited Fluorescence (TPEF) microscopy and spectroscopy, in addition to co- and cross-circularly polarised (CP) reflectance spectrophotometry and Third-Harmonic Generation (THG) spectroscopy, we investigated the optical response of the male \textit{H. coerulea} beetle, unveiling new features such as the optical role of the anisotropy in its photonic structure and its fluorophores' multi-excited states (i.e., states excited by several photons) in the whole range of the visible spectrum. 

One advantage of the two-photon excitation imaging and spectroscopy methods (that are based on a third-order optical process) with respect to one-photon excitation techniques lies in the non-linear intensity-dependent absorption~\cite{Verbiest2009}. Therefore, two-photon absorption is limited to the focal volume within the investigated sample. In theory no photo-induced bleaching occurs outside the focal volume, in contrast with one-photon absorption, which usually gives rise to related strong bleaching in the whole illuminated volume~\cite{Verbiest2009}. In addition, TPEF allows for an extended penetration depth compared to OPEF (from a few tens to hundreds of microns) since the excitation wavelengths are in the near-infrared~\cite{Verbiest2009}, a range for which biological samples are known to be more transparent.

THG is a non-linear process in which three photons combine, through three virtual excited states of the material molecules, into one single photon with a tripled frequency with respect to the incident light (i.e., a third of the incident wavelength). The main advantage of this technique, which has not been fully explored, lies in the possibility to perform label-free analysis both in imaging and spectroscopy~\cite{Rodriguez2017}. 

Due to its recently revealed fluorescent properties and its well-known optical linear response, \textit{H. coerulea}'s photonic structures are a perfect case study for non-linear optical investigations of fluorescent natural photonic architectures in both dry and wet states. TPEF and THG techniques were employed here for the first time in the context of natural photonic structures. The present circularly polarised light and non-linear optical study, together with the steady state and time-resolved fluorescence investigation, reveals the difference between the exited states of the beetle's fluorophores embedded within the photonic structure following one-photon and two-photon absorption as well as the effect of the local environment on their fluorescence emission properties. In addition, they unveil the role of the form anisotropy of the photonic structure inside the beetle's scales, which influences both reflected and emitted light. 

\section{Linear optical and fluorescence properties}
The male \textit{H. coerulea} beetle displays a vivid blue-violet iridescent colour (Figure~\ref{fig:panel1}a) due to a porous multilayer structure (Figure~\ref{fig:panel1}b) located inside the scales covering its elytra and thorax~\cite{Vigneron2005}. The photonic structure is a periodic stack of thin cuticle layers and mixed air-cuticle porous layers (Figure~\ref{fig:panel1}b). Using linear optical and one-photon excited fluorescence techniques, the anisotropy of the beetle's photonic structure (Figure~\ref{fig:panel1}c) is shown to play a role in the reflected CP light, and the fluorophores embedded within this structure are found to possess several excited states, the decay time of which depends on the linear polarisation state of the excitation light.

\begin{figure}
\centering
\includegraphics[width=\columnwidth]{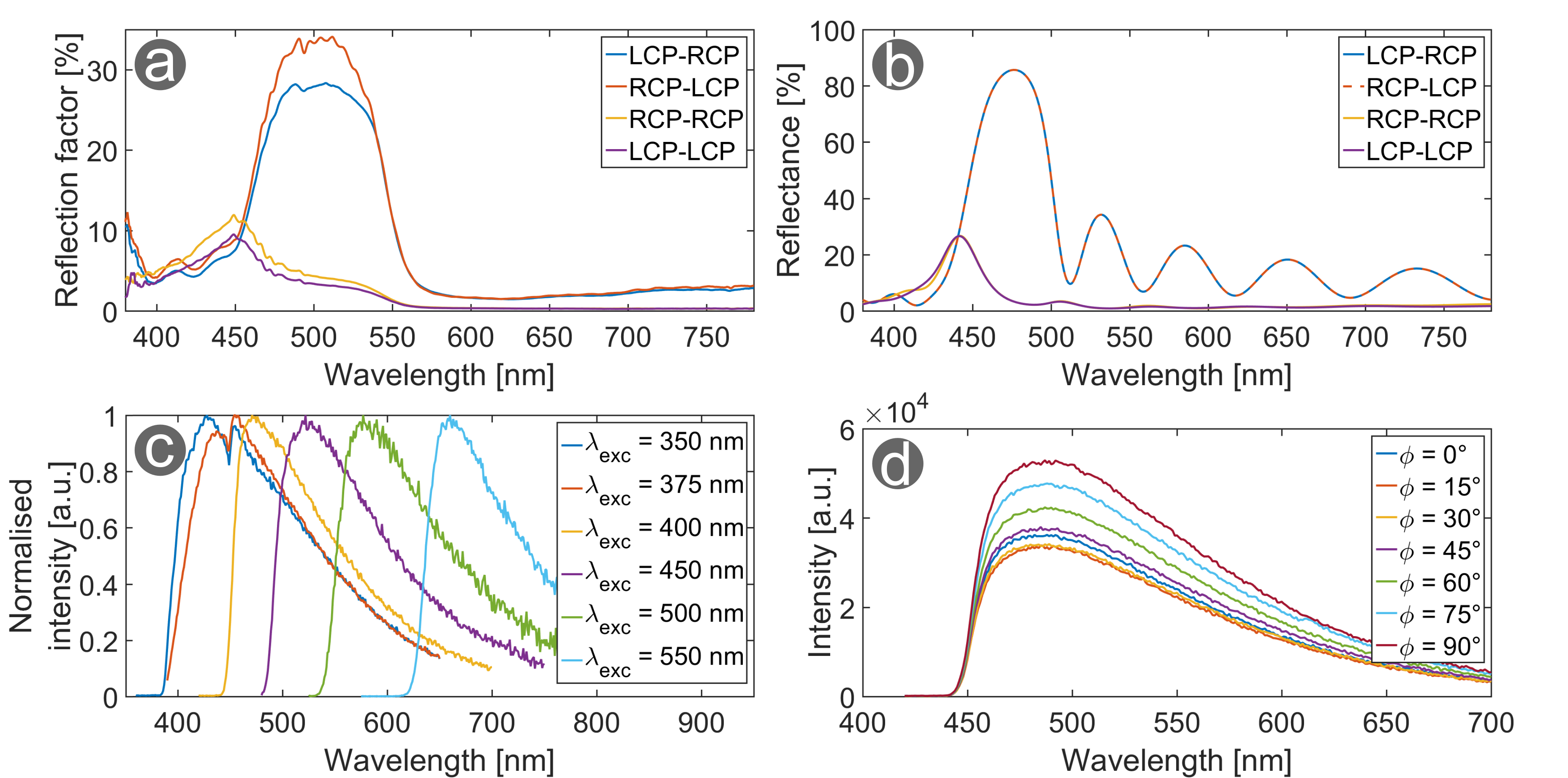}
\caption{\textbf{Linear optical and OPEF responses of \textit{H. coerulea}'s elytra.} (a) In addition to the strong cross-circularly polarised (LCP-RCP and RCP-LCP) reflection at normal incidence, non-negligible co-circularly polarised (LCP-LCP and RCP-RCP) reflection is observed experimentally. (b) The simulated spectra demonstrate that this optical behaviour is due to the presence of form anisotropy in the photonic structure. OPEF emission from embedded fluorophores is significantly modified when tuning (c) the excitation wavelength and (d) the linear polarisation state of the excitation light. $\phi$ is the angle of the rotatable polariser. In (d) the excitation wavelength is equal to 400~nm.}
\label{fig:panel2}
\end{figure}

Previous linear optical studies of the photonic multilayer structure of \textit{H. coerulea} have included measurements of the unpolarised reflectance~\cite{Vigneron2005,Mouchet2017a}, with corresponding simulations that applied isotropic approximations of the structure~\cite{Vigneron2005,VanHooijdonk2012,Mouchet2016c}. In this work, polarised-light techniques are employed in order to identify the influence of anisotropy on the reflectance. The response of the structure to incident circularly polarised (CP) light was measured in terms of CP reflectance components (see Methods section). Both cross-CP and co-CP configurations were included, where cross-CP denotes either left-handed CP (LCP) light reflected from right-handed CP (RCP) incident light or vice versa, and co-CP refers to the incident and reflected CP components having the same handedness. The results (Figure~\ref{fig:panel2}a) show two distinct pairs of spectra. The cross-CP spectra both have a peak centred on 500~nm. A cross-CP polarisation response is similar to that of a planar mirror. Additionally, this structure exhibits conspicuous reflectance spectra in both co-CP configurations, with a smaller peak centred on 450~nm. Such a co-CP reflection response is a feature that has not been extensively observed in scale-bearing insects, and is a strong indication of the presence of anisotropy within the photonic structure. Both intrinsic and form birefringence can be anticipated to confer optical anisotropy upon this structure. Firstly, the cuticular material could have intrinsic birefringence due to the anisotropic self-assembly of the cuticular macromolecules within the multilayer structure~\cite{Berthier2003,Kinoshita2008}. Secondly, the presence of form birefringence~\cite{Xu1996,Tyan1997,Feng2010} is suggested by the morphological characterisation (Figure~\ref{fig:panel1}c), arising due to the rod-like inter-layer spacers forming the mixed air-cuticle layers. The azimuthal directions of these rods typically differ from one such layer to the next, similarly to a woodpile structure~\cite{Joannopoulos2008}.

The CP reflectance of the structure was modelled for normal incidence using a one-dimensional matrix method for anisotropic multilayers~\cite{Ko1998}. The structural dimensions upon which this representative model is based were identified in~\cite{Rassart2009,Mouchet2016c} and are detailed in Methods section. In the absence of direct evidence of intrinsic birefringence for this system, these anisotropic refractive indices were obtained from a calculation of the form birefringence, using the equations given in reference~\cite{Vigneron2005}. Morphological studies~\cite{Vigneron2005,Mouchet2017a} have indicated that the azimuth of the chitin rods is disordered, with variation both from layer to layer and with spatial position within each layer (Figure~\ref{fig:panel1}c). This disorder was accounted for in the model firstly by applying a random azimuth orientation of the in-plane birefringent axes for each period of the structure. Secondly, the result was averaged over multiple simulations, with a different set of random axis orientations applied to the structure in each iteration. The results (Figure~\ref{fig:panel2}b) show that the model reproduces the main features of the experimentally observed CP spectra (Figure~\ref{fig:panel2}a), including the presence of the co-CP signals displaying the same shape. This result and its match with the experimental measurement supports a model of the multilayer in which the air-chitin layers have form anisotropy that contributes to the CP reflectance.

The morphology identified here in \textit{H. coerulea} offers comparison with alternative reflecting structures displaying anisotropy in scale-bearing beetles. The examples of \textit{Lepidiota stigma} and \textit{Cyphochilus} spp. scarab beetles obtain a diffuse white appearance through incoherent scattering using a highly disordered network of chitin rods~\cite{Vukusic2007,Burresi2014,Cortese2015}. There, the anisotropy of the system enhances the brightness by favouring out-of-plane scattering. The case of \textit{H. coerulea} is different since its optical response is angle- and wavelength-dependent due to coherent scattering.

Importantly, the peak width and position for one-photon excited fluorescence emission depends drastically on the excitation wavelength (Figure~\ref{fig:panel2}c). The peak position shifts from 434~nm to 661~nm as the excitation wavelength is increased from 350~nm to 550~nm. These measurements suggest that the embedded fluorophores, the molecular composition of which remains unknown, exhibit complex dynamics with several excited states. In complex organic fluorescent molecules, the presence of several such excited states possibly indicates exciton resonances~\cite{Lakowitz1999,Ramanan2014}. The presence of mixtures of different fluorophores is less likely as this would imply multi-peaked fluorescence emission spectra (at some excitation wavelengths), whereas here we observe single peaks.

In addition, the one-photon excited fluorescence emission spectra depend on the linear polarisation state of the excitation light (Figure~\ref{fig:panel2}d and Supplementary Figure~1). As the rotatable polariser angle $\phi$ is increased from 0° to 90°, the emission intensity increases by a factor 1.55, independent from the excitation wavelength. Due to the weak refractive index contrast between the incident medium (air) and the cuticle material, such an increase in intensity cannot only be explained by a difference of light transmission as a function of the linear polarisation state~\cite{Kinoshita2008}. Furthermore, the radiative decay time $\tau$ is also a function of the excitation light polarisation (Table~1), specifically at emission wavelengths within the photonic band gap of the beetle's scale structure (with a 45° incidence, it ranges from about 410~nm to about 470~nm). For instance, at 456~nm, $\tau$ increases from 6.6~ns to 9.3~ns when $\phi$ is increased from 0° to 90°. At emission wavelengths far from the photonic band gap, $\tau$ is almost constant. Similar but non-polarised measurements reported previously gave rise to much shorter radiative decay times~\cite{Mouchet2016c} ranging from 3.9~ns in the dry state and inside the gap to 1.4~ns in the wet state and outside the gap (Supplementary Table~1). This clearly means that the real excited states and their related radiative decay times are affected by the polarisation state of the excitation light.

The photonic structure anisotropy gives rise to an additional moulding of the reflected light, namely through the selection of reflected light wavelengths depending on the CP light handedness. The embedded fluorophores are found to possess various excited states, the emission spectra and the decays of which are functions of the polarisation states of the excitation light.

\begin {table}
\centering
\begin{tabular}{|c|c|c|c|c|} 
	\hline
    $\lambda_{\text{em}}$ (nm) & $\tau$ (ns) with $\phi$=0° & $\tau$ (ns) with $\phi$=30° & $\tau$ (ns) with $\phi$=60° & $\tau$ (ns) with $\phi$=90°\\
    \hline 
        456 & 6.6 & 8.8 & 8.7 & 9.3\\
    		& 1.4 & 1.3 & 1.3 & 1.4\\
    \hline
    	488 & 9.8 & 10.1 & 8.6 & 8.4\\
    		& 1.2 & 1.2 & 1.1 & 0.9 \\
    \hline
    	520 & 9.8 & 8.6 & 9.8 & 10.0\\
    		& 1.0 & 0.8 & 1.0 & 1.0\\
    \hline
    	580 & 8.8 & 9.2 & 8.9 & 8.7\\
    		& 0.9 & 1.1 & 1.1 & 1.0\\
    \hline
\end{tabular}
\caption{\textbf{Polarisation-dependent radiative and non-radiative decay times of the fluorophores embedded within \textit{H. coerulea}’s elytra.} The radiative decay time $\tau$ (values in top row of each table cell) depends on the light polarisation, while the non-radiative decay time (values in bottom row of each table cell) are constant regardless of the emission wavelength $\lambda_{\text{em}}$ or the angle $\phi$ of the rotatable polariser. Measurements were performed on an elytron in the dry state for $\lambda_{\text{em}}$ located inside (456~nm) and outside (488~nm, 520~nm and 580~nm) the photonic band gap. The incident beam and emitted light collection path formed 45°~angles on either sides of the normal to the sample surface. With such an incidence, the photonic band gap of the structure ranges from about 410~nm to about 470~nm. }
	 \label{tab:table1}
\end {table}

\section{Two-photon excited fluorescence emission}

\begin{figure}
\includegraphics[width=\columnwidth]{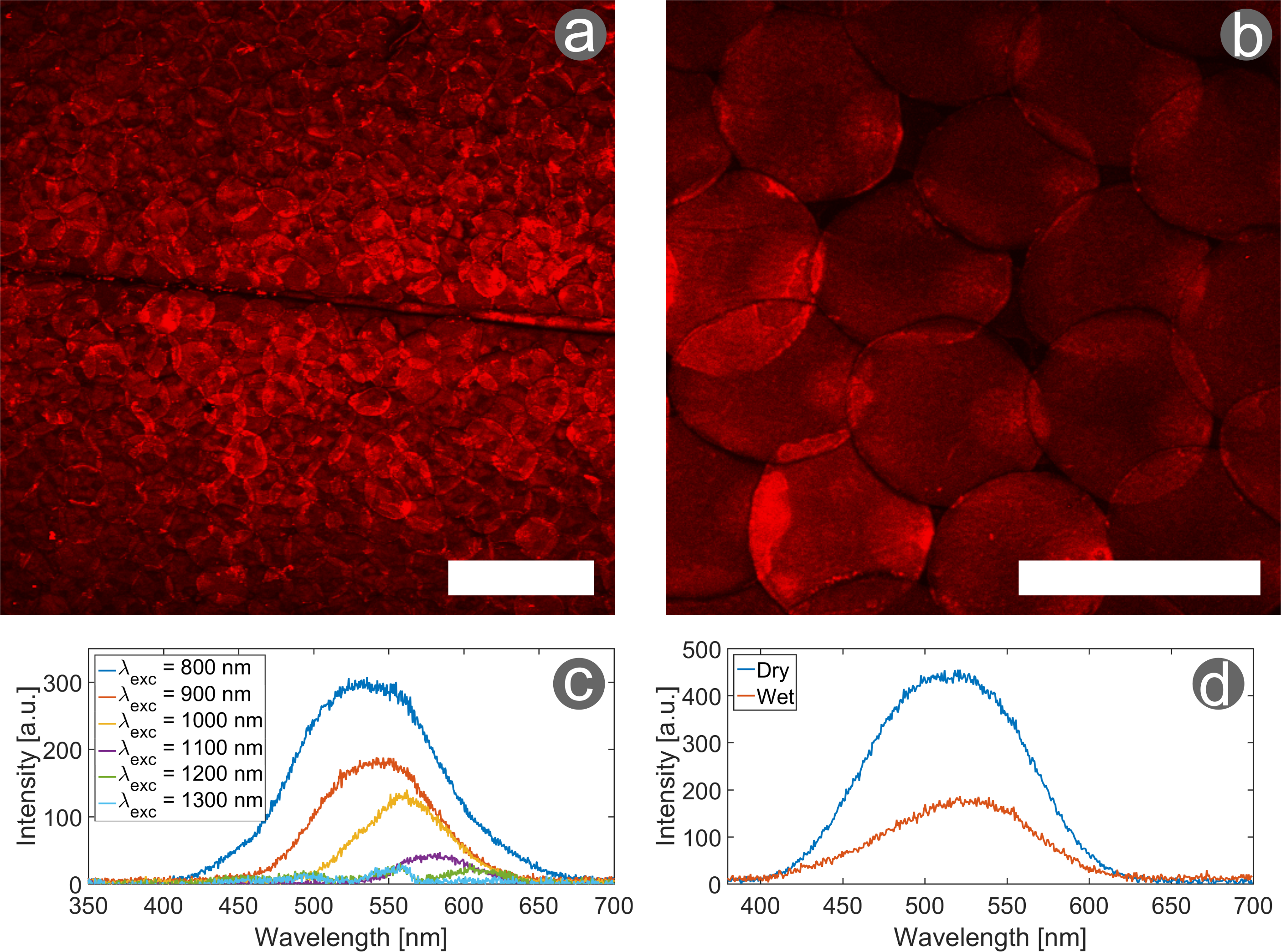}
\caption{\textbf{Two-photon excitation fluorescence response of \textit{H.\ coerulea}'s elytra.} The elytra of the male \textit{H. coerulea} beetle give rise to a strong TPEF response with (a) a 900~nm and (b) a 800~nm excitation light allowing to observe the scales covering the beetle's elytra (with a 75~mW incident power). (c) With a low incident power (20~mW), TPEF excitation spectra can be measured with different excitation wavelengths. (d) Upon contact with water, a small red-shift of the peak wavelength and a decrease in peak intensity is observed with an excitation wavelength equal to 800~nm and an incident power of 20~mW. Scale bars: (a) 200~µm and (b) 100~µm.}
\label{fig:panel3}
\end{figure}

Upon excitation with an incident light beam with a wavelength ranging from 800~nm to 1300~nm (Figure~\ref{fig:panel3}), a TPEF signal is measured from the beetle's scales, even at relatively low incident laser power (20~mW), producing high resolution images (Figure~\ref{fig:panel3}a-b and Supplementary Movie~1). These measurements unveil the presence of multi-excited states of the fluorophores.

\begin{figure}
\centering
\includegraphics[width=0.5\columnwidth]{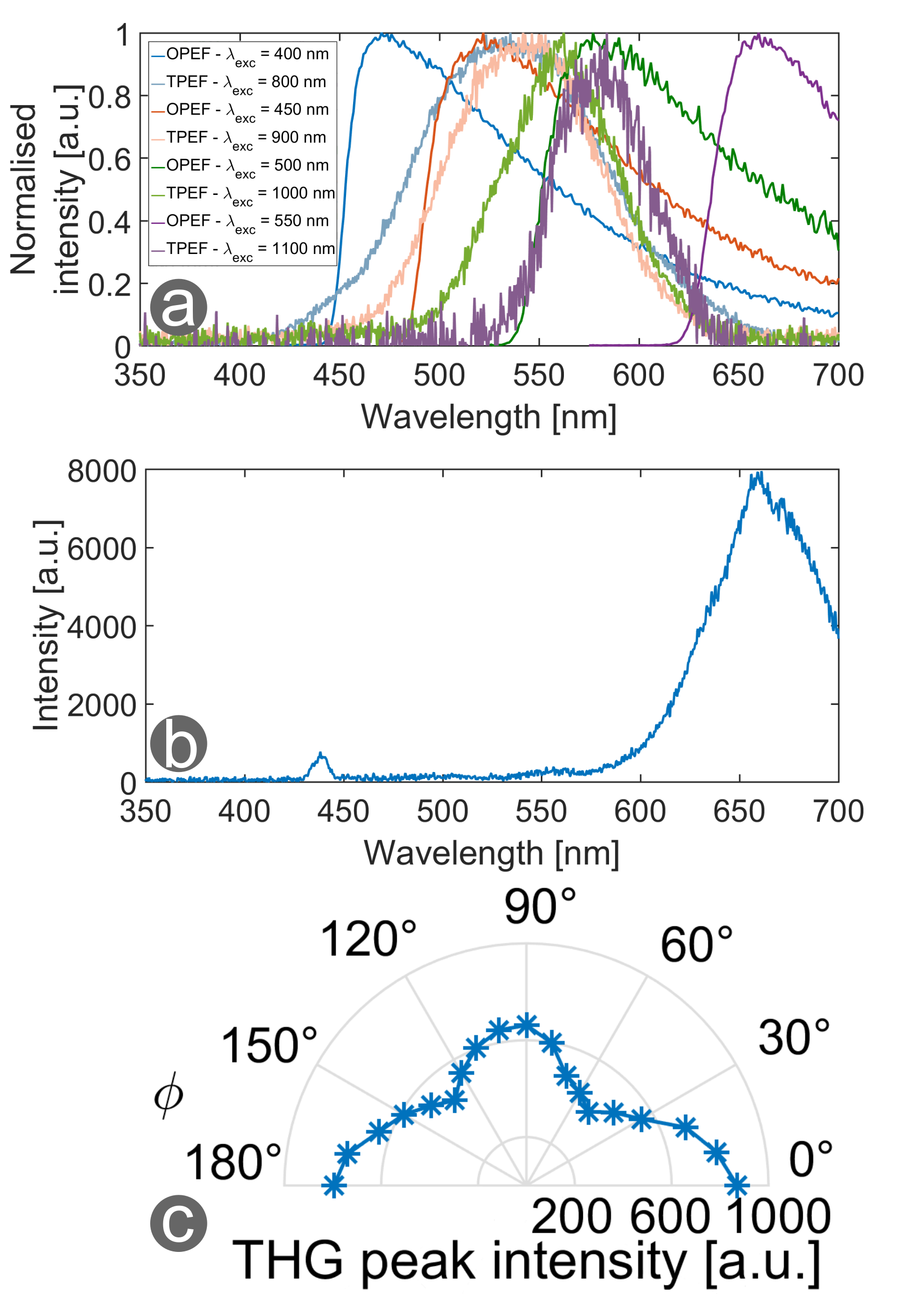}
\caption{\textbf{OPEF, TPEF and THG responses from \textit{H. coerulea}'s elytra.} (a) OPEF and TPEF emission spectra (replotted, respectively, from Figure~\ref{fig:panel2}b and Figure~\ref{fig:panel3}c, after normalisation) are radically different in shape, peak position and FWHM, suggesting that different electronic states are excited in OPEF and TPEF. (b) The increase of the incident power up to 250~mW at 1300~nm excitation wavelength gives rise to a THG signal, the peak of which is located at about 433~nm. (c) The intensity of the THG peak at 433~nm depends on the linear polarisation state of the incident light. $\phi$ is the angle of the rotatable polariser.}
\label{fig:panel4}
\end{figure}

Using 800~nm excitation a TPEF signal with a peak located around 523~nm is detected (Figure~\ref{fig:panel3}c). No second-harmonic generation (SHG) response (expected at 400~nm) could be detected. TPEF emission recorded with different excitation wavelengths (Figure~\ref{fig:panel3}c) confirms the existence of multi-excited states in the fluorophore molecules. Furthermore, the contrasting nature of one- and two-photon absorption processes is expressed in the differences between the TPEF and OPEF spectra in terms of their spectral shapes, peak positions and full widths at half maximum (FWHM) (Figure~\ref{fig:panel4}a). States excited by one-photon absorption with a given wavelength give rise to emission spectra that differ significantly from the states excited by two-photon absorption with twice the corresponding wavelength. For instance, in OPEF and with a 400-nm excitation wavelength, the emission spectrum peaks at 474~nm, while in TPEF and with a 800-nm excitation wavelength, the emission peak is located at 537~nm. In the case of two-photon excitation, some vibrational transitions are additionally enhanced~\cite{Drobizhev2011}, which affect non-radiative relaxation pathways, while in the case of one-photon excitation, simple electronic transitions between real states dominate.

Upon contact of the elytron surface with water, the TPEF response decreases in intensity and red-shifs by about 6~nm (Figure~\ref{fig:panel3}d). Similarly to previously reported OPEF measurements~\cite{Mouchet2016c}, the intensity reduction can be explained by a change in the scattering efficiency at the sample surface due to the presence of water. The peak shift may seem, however, to contrast with the previously reported 17-nm blue-shift of the OPEF peak wavelength (from 463~nm to 446~nm)~\cite{Mouchet2016c}. In the dry state, with a 355~nm excitation wavelength, the emission peak was positioned within the photonic band gap, a range where the local density of optical state is very sensitive to changes in the refractive index contrast induced upon contact with water~\cite{Mouchet2016c}. This opposite and smaller shift observed with TPEF can be explained by the fact that the selection rules are entirely different for the one- and two-photon absorption processes~\cite{Drobizhev2011}. The corresponding excited states are different, as observed in Figure~\ref{fig:panel4}a, and, therefore, their radiative and non-radiative relaxations are different in the presence of liquid. Additionally and in contrast with the OPEF emission peak, the TPEF emission peak is far from the photonic band gap, i.e., in a spectral range where the local density of optical states is not much affected by the presence of liquid in the pores of the structure~\cite{Mouchet2016c}. All together this produces a 6-nm red-shift of TPEF and a 17-nm blue-shift of OPEF. Finally, the liquid-induced red- (blue-)shift of TPEF (OPEF) suggests that the polarities of the excited states involved in one- and two-photon absorption processes are different. In the TPEF case, the presence of water modifies slightly the environment around the ground and real excited states making them slightly more polar in the wet sample than in the dry sample, resulting in a decrease of the energy difference between excited and ground states~\cite{Lakowicz2006}, as indicated by the red-shift of TPEF upon contact with water.

These observations of the TPEF signal have highlighted the multi-excited states nature of the fluorescent molecules. Upon contact with water, the difference between the measured red-shift of the TPEF peak and the blue-shift of OPEF may be explained by the emission wavelength band that is far from the photonic band gap, by the difference in selection rules for one- and two-photon absorption processes and by the possible difference in polarities of the excited states.

\section{Third-Harmonic Generation}

Using THG spectroscopy, the beetle's elytra are found to give rise to a polarisation-dependent signal which represents an additional consequence of the photonic structure's form anisotropy.

In spite of the absence of a detectable SHG signal at 20-mW incident power (Figure~\ref{fig:panel3}c), a THG signal with a peak around 433~nm was measured using a higher incident power (250~mW) and an excitation (fundamental) wavelength equal to 1300~nm (Figure~\ref{fig:panel4}b). The peak positioned around 650~nm is mainly due to TPEF signal, which may possibly obscure a much weaker SHG signal. The intensity of the THG peak depends on the polarisation state of the incident light (Figure~\ref{fig:panel4}c). 
THG can arise from any material regardless of the symmetry exhibited by the constituting molecules. In addition to the chemical composition of \textit{H. coerulea}'s scales and the structure of molecules that form this biological material (which are far from perfectly known~\cite{Mouchet2016a}, like for any other beetle), the non-linear optical response from the beetle's elytra is likely to be affected by light interference within the host material, i.e.,\ by the scale's photonic structure. The polarisation-dependent response could be attributed to the form anisotropy of the material, highlighted by the co- and cross-CP reflectance measurements (Figure~\ref{fig:panel2}a,b), since the THG intensity of an isotropic material does not depend on the incident light polarisation~\cite{Rodriguez2017,Chen2014,Wu2016,Cui2017}.

\section{Conclusion}

Beyond its structural blue-violet colouration, the male \textit{H. coerulea} beetle is an instructive example of fluorescence emission controlled by natural photonic structures. Using fluorescence, linear and non-linear optical techniques, we have gained a better understanding of the beetle's elytra optical properties and of the excited states of the embedded fluorophores. Observations of a non vanishing reflection in co-circularly polarised light configuration, as well as of a polarisation-dependent THG signal, unveil the optical role of form anisotropy, potentially responsible for an additional control of the fluorescence emission. OPEF and TPEF observations show that one single type of fluorescent molecules having several excited and multi-excited states is present in the scales' biological tissues. The corresponding one-photon and two-photon absorption excited states were observed to exhibit very different emission spectra. A spectral red-shift of the TPEF peak induced upon contact of the elytra with water suggests that the multi-excited states may have different polarities. This seminal investigation is believed to promote the study of photonic structures in natural organisms using multi-photon excited fluorescence and non-linear optical techniques in order to derive a comprehensive picture of their fluorescent and optical properties. This novel approach will indubitably result in a deeper questioning of the biological roles of fluorescence and other optical properties of living organisms.

\section{Methods}
\subsection{Beetle samples}
{Male \textit{H. coerulea} specimens were collected in Saumane and Auzon (France) in June 2014. No further sample preparation was necessary. All the analyses were performed on the beetle's elytra.

\subsection{Co- and cross-circularly polarised reflectance spectrophotometry}
The beetle elytron was characterised using co- and cross-circularly polarised reflectance spectrophotometry. The experimental set-up was described in a previous work~\cite{Finlayson2017} and consisted of an Ocean Optics (Delray Beach, FL, USA) HPX-2000 broadband fibre-coupled light source and an Ocean Optics USB2000+ spectrometer. CP light beam was generated by a rotatable polariser and an achromatic Fresnel rhomb quarter-wave retarder, orientated at 45° azimuth. The CP light handedness was chosen by orientating the polariser azimuth at either 0° or 90°. The incident light was focused onto a single beetle elytron at normal incidence using an achromatic ×10 objective lens, producing a beam spot diameter of approximately 30 µm. This lens also collected light reflected by the sample, which was routed by a beam-splitter through a CP analyser comprising the Fresnel rhomb and a second rotatable polariser. The calibration was performed using a plane aluminium mirror.

\subsection{Optical modelling}
The CP reflectance of the structure was modelled for normal incidence using a one-dimensional $4\times4$ matrix method for anisotropic multilayers~\cite{Ko1998}. The structural dimensions upon which this representative model is based were identified in~\cite{Rassart2009,Mouchet2016c} and are as follows. The multilayer contained 12 bilayer periods, each comprising a 35~nm-thick chitin layer (isotropic, $n = 1.56$) and a 140~nm-thick air-chitin layer. The chitin was assumed to be lossless. The properties of the air-chitin layers were based on a laterally-periodic assembly of rectangular chitin rods of width 90~nm, spaced by air gaps of 85~nm width. The air-chitin layers were modelled with form-birefringent refractive indices $n_o = 1.32$ and $n_e = 1.20$, which were calculated using the equations given in reference~\cite{Vigneron2005}. A 100~nm-thick cap layer of chitin (isotropic, $n = 1.56$) was applied, which represented the enclosing permeable envelope~\cite{Mouchet2016a}. Disorder in the azimuthal orientation of the chitin rods was accounted for in the model firstly by applying a random azimuth orientation of the in-plane birefringent axes for each period of the structure. The result was averaged over 50 iterations, with a different set of random axis orientations applied to the structure in each iteration. 

\subsection{Spectrofluorimetry}
Fluorescence measurements were performed using an Edinburgh Instruments (Livingston, UK) FLSP920 UV–vis–NIR spectrofluorimeter equipped with a Hamamatsu (Hamamatsu City, Japan) R928P photomultiplier tube. Excitation and emission spectra were recorded using a 450-W xenon lamp as the steady state excitation source. The incident light was formed at 45° angle with the direction normal to the surface of sample and the emitted light was detected at a 45° angle on the other side of the normal direction. The sample was excited at different excitation wavelengths: 400~nm, 450~nm, 500~nm and 550~nm. Time-resolved dynamics of fluorescence emission from the sample were recorded using an µF920H xenon Flashlamp light source, operating at a frequency of 100 Hz, and with pulse width of 1~µs. The same excitation wavelength as for the steady state measurements were used. The emission dynamics were observed at the maximum of the emission peak. The recorded time-resolved dynamics were fitted to double exponential functions, with coefficients of determination $R^{2}$ ranging from 0.985 to 0.990 for all fits. The polarisation steady state and time-resolved measurements were recorded on the same spectrofluorimeter which is equipped with Glan-Thompson polarisers. The polarisation state of excitation light for both measurements was selected by inserting a polariser in the beam path before the sample. The polariser was computer-controlled and the angles of light polarisation could be varied between 0° and 90°. The excitation light polarisation angle was tuned by steps of 15°. The unpolarised emission spectra were corrected for detector sensitivity. 

\subsection{Multiphoton microscopy and spectrometry}
The multiphoton microscopy images were taken with an Olympus (M\"{u}nster, Germany) BX61 WI-FV1200-M system. A Spectra-Physics (Santa Clara, CA, USA) InSight DS+ laser (82-MHz repetition rate, 120-fs pulse width, p-polarised) was used for excitation, at 800~nm, 900~nm and 1000~nm fundamental wavelengths. To regulate the intensity of the laser beam, an achromatic half-wave plate and an s-directed polariser were placed directly behind the laser, which resulted thus in a s-polarised laser beam, which was used for all experiments. A 15X LMV objective (NA 0.3) and 50X SLMPlan N (NA 0.35) were used for magnification, signal detection happened non-descanned in backwards reflection through a Hamamatsu R3896 photomultiplier tube. Three detection cubes were used, each for a different fundamental wavelength. For 800~nm, the SHG and TPEF signals were divided through a 425~LPXR dichroic mirror. The SHG-signal was further filtered with a 405/10 bandpass filter. Similar detection cubes were used with a 470~LPXR dichroic mirror and 450/7x bandpass filter (900~nm) and a 525~LPXR dichroic mirror and 500/20m bandpass filter (1000~nm). The signal at each pixel is depicted in the images as different intensities in a false red colour. Due to the elytra curvature and roughness, an extended depth of focus function was used in order to obtain in-focus microscopic images. No bleaching was observed, as the fluorescence signal stayed stable over time.

The spectral measurements were performed using the same laser at a 45° incidence angle, with different excitation wavelengths: 800~nm, 900~nm, 1000~nm, 1100~nm, 1200~nm and 1300~nm. An achromatic lens (60~mm EFL) was placed in front of the beetle's elytron in order to focus the laser beam. The spectra were recorded in reflection and collected by an achromatic aspheric condenser lens (focal length 30~mm). The beam was focused on the entrance slit of a Bruker (Billerica, MA, USA) 500~is/sm spectrograph, and then recorded by the Andor Solis (Belfast, UK) iXon Ultra 897 EMCCD camera. In order to eliminate laser light entering from the detector, additional SCHOTT (Mainz, Germany) BG39 and KG5 filters were positioned in the beam path in front of the sample. Additionally, a long-pass filter was placed in front of the first lens to remove the higher-order harmonics of the laser itself. The linear polarisation THG measurements were performed by placing and rotating a second half-wave plate in the beam path, behind the polariser. The 0° and 90° angles of light polarisation corresponded respectively to s- and p-polarisations. After subtracting the background, the THG peak was fitted to a Gaussian curve and the surface of the THG peak was calculated. This value was plotted in function of the polarisation of the incident light beam.

\textbf{Acknowledgements}
The authors thank Louis Dellieu (Department of Physics, UNamur) for technical support during the collection of samples. S.R.M. was supported by Wallonia-Brussels International (WBI) through a Postdoctoral Fellowship for Excellence program WBI.WORLD and by the Belgian National Fund for Scientific Research (F.R.S.-FNRS) as a Postdoctotal Researcher. S.V.C. is grateful to FWO Flanders for his postdoctoral fellowship. R.V.D. thanks the Hercules Foundation (project AUGE/09/024 "Advanced Luminescence Setup") for funding. T.V. acknowledges financial support from the Hercules Foundation. B.K. acknowledges financial support from the "Action de Recherche Concert\'{e}e" (BIOSTRUCT project‚ No.10/15-033) of UNamur, from Nanoscale Quantum Optics COST-MP1403 action and from F.R.S.-FNRS; Interuniversity Attraction Pole: Photonics@be (P7-35, Belgian Science Policy Office).

\textbf{Author contributions}
B.K. and S.R.M. conceived the original project. E.D.F. and S.R.M. conducted the co- and cross-CP reflectance spectrophotometry. E.D.F. performed the optical modelling. D.M. and B.K. performed the OPEF measurements. C.V. and S.V.C. conducted the TPEF and THG measurements. B.K., S.R.M. and C.V. analysed the related data. S.R.M., C.V., D.M., S.V.C., E.D.F., R.V.D., O.D., T.V., B.M., P.V. and B.K. discussed the results. S.R.M. and B.K. wrote the manuscript with input from C.V., D.M. and E.D.F. All authors commented on the manuscript and gave approval to the final version of the manuscript.

\textbf{Additional information}
Supplementary information is available in the online version of the paper. Reprints and permissions information is available online at www.nature.com/reprints. Correspondence and requests for materials should be addressed to S.R.M. and B.K.

\textbf{Competing financial interests}
The authors declare no competing financial interests.


\begin{thebibliography}{99}
\bibitem{Pavan1954} M. Pavan and M. Vachon, "Sur l'existence d'une substance fluorescente dans les téguments des Scorpions (Arachnides)," C. R. Acad. Sci. 239, 1700-1702, 1954.
\bibitem{Tani2004} K. Tani, F. Watari, M. Uo and M. Morita, "Fluorescent Properties of Porcelain-Restored Teeth and Their Discrimination," Mater. Trans. 45, 1010-1014, 2004.
\bibitem{Vukusic2005} P. Vukusic and I. Hooper, "Directionally Controlled Fluorescence Emission in Butterflies," Science 310, 1151, 2005.
\bibitem{Welch2012} V. L. Welch, E. Van Hooijdonk, N. Intrater and J.-P. Vigneron, "Fluorescence in insects," Proc. SPIE 8480, 848004, 2012.
\bibitem{Lagorio2015} M. G. Lagorio, G. B. Cordon and A. Iriel, "Reviewing the relevance of fluorescence in biological systems," Photochem. Photobiol. Sci. 14, 1538-1559, 2015.
\bibitem{Marshall2017} J. Marshall and S. Johnsen, "Fluorescence as a means of colour signal enhancement," Phil. Trans. R. Soc. B 372, 20160335, 2017.
\bibitem{Taboada2017} C. Taboada, A. E. Brunetti, F. N. Pedron, F. Carnevale Neto, D. A. Estrin, S. E. Bari, L. B. Chemes, N. Peporine Lopes, M. G. Lagorio and J. Faivovich, "Naturally occurring fluorescence in frogs," Proc. Natl. Acad. Sci. USA 114, 3672-3677, 2017.
\bibitem{Vigneron2005} J.-P. Vigneron, J.-F. Colomer, N. Vigneron and V. Lousse, "Natural layer-by-layer photonic structure in the squamae of \textit{Hoplia coerulea} (Coleoptera)," Phys. Rev. E 72, 061904, 2005.
\bibitem{Rassart2009} M. Rassart, P. Simonis, A. Bay, O. Deparis and J.- P. Vigneron, "Scale coloration change following water absorption in the beetle \textit{Hoplia coerulea} (Coleoptera)," Phys. Rev. E 80, 31910, 2009.
\bibitem{Mouchet2016b} S. R. Mouchet, T. Tabarrant, S. Lucas, B. L. Su, P. Vukusic and O. Deparis, "Vapor sensing with a natural photonic cell," Opt. Exp. 24, 12267-12280, 2016.
\bibitem{Mouchet2016a} S. R. Mouchet, E. Van Hooijdonk, V. L. Welch P. Louette, J.-F. Colomer, B.-L. Su and O. Deparis, "Liquid-induced colour change in a beetle: the concept of a photonic cell," Sci. Rep. 6, 19322, 2016.
\bibitem{Mouchet2017b} S. R. Mouchet, E. Van Hooijdonk, V. L. Welch, P. Louette, T. Tabarrant, P. Vukusic, S. Lucas, J.-F. Colomer, B.-L. Su and O. Deparis, "Assessment of environmental spectral ellipsometry for characterising fluid-induced colour changes in natural photonic structures," Mater. Today Proc. 4, 4987-4997, 2017.
\bibitem{VanHooijdonk2012} E. Van Hooijdonk, S. Berthier and J.-P. Vigneron, "Bio-inspired approach of the fluorescence emission properties in the scarabaeid beetle \textit{Hoplia coerulea} (Coleoptera): Modeling by transfer matrix optical simulations," J. Appl. Phys. 112, 114702, 2012.
\bibitem{Mouchet2016c} S. R. Mouchet, M. Lobet, B. Kolaric, A. M. Kaczmarek, R. Van Deun, P. Vukusic, O. Deparis and E. Van Hooijdonk, "Controlled fluorescence in a beetle's photonic structure and its sensitivity to environmentally induced changes," Proc. R. Soc. London B Biol. Sci. 283, 20162334, 2016.
\bibitem{Mouchet2017a} S. R. Mouchet, M. Lobet, B. Kolaric, A. M. Kaczmarek, R. Van Deun, P. Vukusic, O. Deparis and E. Van Hooijdonk, "Photonic scales of \textit{Hoplia coerulea} beetle: any colour you like," Mater. Today Proc. 4, 4979-4986, 2017.
\bibitem{Kolaric2007} B. Kolaric, K. Baert, M. Van der Auweraer, R. A. L. Vallée and K. Clays, "Controlling the fluorescence resonant energy transfer by photonic crystal band gap engineering," Chem. Mater 19, 5547-5552, 2007.
\bibitem{Gonzalez2012} L. González-Urbina, K. Baert, B. Kolaric, J. Pérez-Moreno and K. Clays, "Linear and nonlinear optical properties of colloidal photonic crystals," Chem. Rev. 112, 2268-2285, 2012.
\bibitem{Rowlands2017} C. J. Rowlands, D. Park, O. T. Bruns, K. D. Piatkevich, D. Fukumura, R. K. Jain, M. G. Bawendi, E. S. Boyden and P. T. C. So, "Wide-field three-photon excitation in biological samples," Light Sci. Appl. 6, e16255, 2017.
\bibitem{Verbiest2009} T. Verbiest, K. Clays and V. Rodriguez, "Second-Order Nonlinear Optical Characterization Techniques - An Introduction," CRC Press Taylor \& Francis Group, Boca Raton, FL, 2009.
\bibitem{Rodriguez2017} V. Rodriguez, "Polarization-Resolved Third-Harmonic Scattering in Liquids," J. Phys. Chem. C 121, 8510-8514, 2017.
\bibitem{Berthier2003} S. Berthier, "Iridescences, les couleurs physiques des Insectes," Springer, Paris, 2003.
\bibitem{Kinoshita2008} S. Kinoshita, "Structural Colors in the Realm of Nature," World Scientic Publishing Co, Singapore, 2008.
\bibitem{Xu1996} F. Xu, R.-C. Tyan, P.-C. Sun, Y. Fainman, C.-C. Cheng and A. Scherer, "Form-birefringent computer-generated holograms," Opt. Lett. 21, 1513-1515, 1996.
\bibitem{Tyan1997} R.-C. Tyan, A. A. Salvekar, H.-P. Chou, C.-C. Cheng, A. Scherer, P.-C. Sun, F. Xu and Y. Fainman, "Design, fabrication, and characterization of form-birefringent multilayer polarizing beam splitter," J. Opt. Soc. Am. A 14, 1627-1636, 1997.
\bibitem{Feng2010} L. Feng, Z. Liu, V. Lomakin and Y. Fainman, "Form birefringence metal and its plasmonic anisotropy," Appl. Phys. Lett. 96, 041112, 2010.
\bibitem{Joannopoulos2008} J. D. Joannopoulos, S. G. Johnson, J. N. Winn and R. D. Meade, "Photonic Crystals: Molding the Flow of Light," Princeton University Press, Princeton, 2008.
\bibitem{Ko1998} D. Y. K. Ko and J. R. Sambles, "Scattering matrix method for propagation in stratified media: attenuated total reflection studies of liquid crystals," J. Opt. Soc. Am. A 5, 1863–1866, 1988.
\bibitem{Vukusic2007} P. Vukusic, B. Hallam and J. Noyes, "Brilliant Whiteness in Ultrathin Beetle Scales," Science 315, 2007.
\bibitem{Burresi2014} M. Burresi, L. Cortese, L. Pattelli, M. Kolle, P. Vukusic, D. S. Wiersma, U. Steiner and S. Vignolini, "Bright-White Beetle Scales Optimise Multiple Scattering of Light," Sci. Rep. 4, 6075, 2014. 
\bibitem{Cortese2015} L. Cortese, L. Pattelli, F. Utel, S. Vignolini, M. Burresi and D. S. Wiersma, "Anisotropic Light Transport in White Beetle Scales," Adv. Optical Mater. 3, 1337-1341, 2015.
\bibitem{Lakowitz1999} J. R. Lakowitz, "Principles of Fluorescence Spectroscopy," Kluwer Academic/Plenum Publishers, New York, NY, 1999.
\bibitem{Ramanan2014} C. Ramanan, C. Hoon Kim, T. J. Marks and M. R. Wasielewski, "Excitation Energy Transfer within Covalent Tetrahedral Perylenediimide Tetramers and Their Intermolecular Aggregates," J. Phys. Chem. C 118, 16941–16950, 2014.
\bibitem{Drobizhev2011} M. Drobizhev, N. S. Makarov, S. E. Tillo, T. E. Hughes and A. Rebane, "Two-photon absorption properties of fluorescent proteins," Nat. Meth. 8, 393-399, 2011.
\bibitem{Lakowicz2006} J. R. Lakowicz, "Principles of Fluorescence Spectroscopy," 3rd ed., Springer, New York, 187-215, 2006.
\bibitem{Chen2014} S. Chen, G. Li, F. Zeuner, W. H. Wong, E. Y. B. Pun, T. Zentgraf, K. W. Cheah and S. Zhang, "Symmetry-Selective Third-Harmonic Generation from Plasmonic Metacrystals," Phys. Rev. Lett. 113, 033901, 2014.
\bibitem{Wu2016} H.-Y. Wu, Y. Yen and C.-H. Liu, "Observation of polarization and thickness dependent third-harmonic generation in multilayer black phosphorus," Appl. Phys. Lett. 109, 261902, 2016.
\bibitem{Cui2017} Q. Cui, R. A. Muniz, J. E. Sipe and H. Zhao, "Strong and anisotropic third-harmonic generation in monolayer and multilayer ReS$_2$," Phys. Rev. B 95, 165406, 2017.
\bibitem{Finlayson2017} E. D. Finlayson, L. T. McDonald and P. Vukusic, "Optically ambidextrous circularly polarised reflection from the chiral cuticle of the scarab beetle \textit{Chrysina resplendens}," J. R. Soc. Interface 14, 20170129, 2017.
\end{thebibliography}
\end{document}